\documentclass[preprint,11pt]{elsarticle}
\usepackage{amsmath}
\usepackage{epsfig}
\usepackage{epstopdf}
\usepackage{subcaption}
\usepackage{graphicx}
\biboptions{numbers,sort&compress}
\usepackage{hyperref}
\hypersetup{
	colorlinks   = true, 
	urlcolor     = blue, 
	linkcolor    = blue, 
	citecolor   = blue 
}

\usepackage{amssymb}

\journal{Nuclear Physics A}

\begin{document}

\begin{frontmatter}



\title{HARTREE-FOCK-BOGOLIUBOV CALCULATION OF GROUND STATE PROPERTIES OF EVEN-EVEN AND ODD Mo AND Ru ISOTOPES}
 \author{Y. EL BASSEM}
 \author{M. OULNE\corref{cor1}}
 \ead{oulne@uca.ma}
 \cortext[cor1]{Correspondant author}
 \address{High Energy Physics and Astrophysics Laboratory, Department of Physics, \\Faculty of Sciences SEMLALIA, Cadi Ayyad University,  \\P.O.B. 2390, Marrakesh, Morocco.}

\begin{abstract}
In a previous work [\href{http://www.worldscientific.com/doi/abs/10.1142/S0218301315500731}{Int. J. Mod. Phys. E 24, 1550073 (2015)}], hereafter referred as paper\,I, we have investigated the ground-state properties of Nd, Ce and Sm isotopes within Hartree-Fock-Bogoliubov method with SLy5 skyrme force in which the pairing strength has been generalized with a new proposed formula. However, that formula is more appropriate for the region of Nd. In this work, we have studied the ground-state properties of both even-even and odd Mo and Ru isotopes. For this, we have used Hartree-Fock-Bogoliubov method with SLy4 skyrme force, and a new formula of the pairing strength which is more accurate for this region of nuclei. The results have been compared with available experimental data, the results of Hartree-Fock-Bogoliubov calculations based on the D1S Gogny effective nucleon-nucleon interaction and predictions of some nuclear models such as Finite Range Droplet Model (FRDM) and Relativistic Mean Field (RMF) theory.
\end{abstract}


\end{frontmatter}


\section{Introduction}
One of the major aims of research in nuclear physics is to make reliable predictions with one nuclear model in order to describe the ground-state properties of all nuclei in the nucleic chart. Several approaches have been developed to study ground-state and single-particle (s.p) excited states properties of even-even and odd nuclei.
Due to the lack of fully understanding the strong interaction and to the numerical difficulties in handling the nuclear many-body problem, non-relativistic \cite{Terasaki,Dobaczewski,Chabanat,Stoitsov2000,Teran} and relativistic \cite{Ring96,Lalazissis} mean field theories have received much attention for describing the ground-state properties of nuclei. One of the most important phenomenological approaches widely used in nuclear structure calculations is the Hartree-Fock-Bogoliubov method \cite{Yamagami}, which unifies the self-consistent description of nuclear orbitals, as given by Hartree-Fock (HF) approach, and the Bardeen-Cooper-Schrieffer (BCS) pairing theory \cite{Bardeen} into a single variational theory.

Molybdenum (Mo, Z=42) and Ruthenium(Ru, Z=44), as well as all nuclei which have neutron numbers close to the magic number N=50, exhibit many interesting nuclear properties such as anomalous behavior in the isotope shifts and large changes of shape \cite{Lalazissis95,Rodriguez}.

In this paper we are interested in calculating and analyzing some ground-state properties of even-even and odd $Mo$ isotopes for a wide range of neutron numbers, by using Skyrme-Hartree-Fock-Bogoliubov method and a new generalized formula for the pairing strength. The ground-state properties we have focused on are binding energy, one- and two-neutron separation energies, charge, proton and neutron radii, neutron pairing gap and quadrupole deformation. We have also performed similar calculations for $Ru$ which is in the surroundings of $Mo$.

The paper is organized in the following way : In Section 2, we briefly present the Hartree-Fock-bogoliubov method. Some details about the numerical calculations are presented in Section 3, while in Section 4, we present our results and discussion. A conclusion is given in Section 5.

\section{Hartree-Fock-Bogoliubov Method}
The Hartree-Fock-Bogoliubov (HFB) \cite{Bogoliubov,Ring} framework has been extensively discussed in the literature \cite{Ring,Dobaczewski84,Dobaczewski96,Bender} and will be briefly introduced here.\\
In HFB method, a two-body Hamiltonian of a system of fermions can be expressed in terms of a set of annihilation and creation
operators $(c, c^\dagger)$:
\begin{equation}
H=\sum_{n_1 n_2} e_{n_1 n_2} c_{n_1}^\dagger c_{n_2} + \frac{1}{4} \sum_{n_1 n_2 n_3 n_4} \bar{\nu}_{n_1 n_2 n_3 n_4} c_{n_1}^\dagger c_{n_2}^\dagger c_{n_4} c_{n_3}
\label{eq1}
\end{equation}
with the first term corresponding to the kinetic energy and $\bar{\nu}_{n_1 n_2 n_3 n_4}=\langle n_1 n_2 | V | n_3 n_4 - n_4 n_3 \rangle$ are anti-symmetrized two-body interaction matrix-elements. \vspace{0.4em}
So, the ground-state wave function $|\varPhi\rangle$ is defined as the quasi-particle vacuum $\alpha_k|\varPhi\rangle=0$, in which the quasi-particle operators $(\alpha,\alpha^\dagger)$ are connected to the original particle ones via a linear Bogoliubov transformation :
\begin{equation}
\alpha_k=\sum_n (U_{nk}^* c_n + V_{nk}^* c_n^\dagger),~~~~~~~~~~~~\alpha_k^\dagger=\sum_n (V_{nk} c_n + U_{nk} c_n^\dagger),
\end{equation}
which can be rewritten in the matrix form as :
\begin{eqnarray}
\left(\begin{array}{c} \alpha \\ \alpha^\dagger \end{array}\right)=\left(\begin{array}{c  c} U^\dagger & V^\dagger \\
V^T & U^T
\end{array} \right)\left(\begin{array}{c} c \\ c^\dagger \end{array} \right) , \, \label{eqhfb}
\end{eqnarray}
The matrices U and V satisfy the relations:
\begin{equation}
U^\dagger U+V^\dagger V=1,~~~~U U^\dagger + V^* V^T=1,~~~~U^T V+V^T U =0,~~~~U V^\dagger+ V^* U^T=0.
\end{equation}

In terms of the normal $\rho$ and pairing $\kappa$ one-body density matrices, defined as :
\begin{equation}
\rho_{nn'}=\langle\Phi|c_{n'}^\dagger c_{n}|\Phi\rangle=(V^{*}V^{T})_{nn'}, \hspace{0.3 in}
\kappa_{nn'}=\langle\Phi|c_{n'}c_{n}|\Phi\rangle=(V^{*}U^{T})_{nn'}\,,
\label{matrices}
\end{equation}
the expectation value of the Hamiltonian (\ref{eq1}) is expressed as an energy functional
\begin{equation}
E[\rho,\kappa]=\frac{\langle\Phi|H|\Phi\rangle}{\langle\Phi|\Phi\rangle}=
\textrm{Tr}[(e+\frac{1}{2}\Gamma)\rho]-\frac{1}{2}\textrm{Tr}[\Delta\kappa^{*}]
\label{energyfunctional}
\end{equation}
where\\
\begin{equation}
\Gamma_{n_{1}n_{3}}=\sum_{n_{2}n{4}}\bar\upsilon_{n_{1}n_{2}n_{3}n_{4}}\rho_{n_{4}n_{2}}\,,~~~~~~~~~~         \Delta_{n_{1}n_{2}}=\frac{1}{2}\sum_{n_{3}n{4}}\bar\upsilon_{n_{1}n_{2}n_{3}n_{4}}\kappa_{n_{3}n_{4}}\,.
\end{equation}
The variation of the energy (\ref{energyfunctional}) with respect to $\rho$ and $\kappa$ leads to the HFB equations :
\begin{eqnarray}
\left(\begin{array}{cc} e+\Gamma-\lambda & \Delta \\
-\Delta^* & -(e+\Gamma)^*+\lambda
\end{array} \right)\left(\begin{array}{c} U \\ V \end{array} \right) =E\left(\begin{array}{c} U \\ V \end{array}\right), \, \label{eqhfb}
\end{eqnarray}
where $\Delta$ and $\lambda$ denote the pairing potential and Lagrange multiplier, introduced to fix the correct average particle number, respectively.\\
It should be stressed that the energy functional (\ref{energyfunctional}) contains terms that cannot be simply related to some prescribed effective interaction \cite{Bender}. 
In terms of Skyrme forces, the HFB energy (\ref{energyfunctional})
has the form of local energy density functional :

\begin{equation}
E[\rho,\tilde{\rho}]=\int d^{3}\textrm{H}(\textbf{r}),
\label{skyrmeefunctional}
\end{equation}

where \textrm{H}(\textbf{r})
is the sum of the mean field and pairing energy densities. The variation of
the energy (\ref{skyrmeefunctional}) according to the particle local
density $\rho$ and pairing local density $\tilde{\rho}$ results in the
Skyrme HFB equations :

\begin{eqnarray}
\begin{split}
\sum_{\sigma^{\prime}}\left(\begin{array}{cc} h(\textbf{r},\sigma,\sigma^{\prime}) & \Delta(\textbf{r},\sigma,\sigma^{\prime}) \\
\Delta(\textbf{r},\sigma,\sigma^{\prime}) & -h(\textbf{r},\sigma,\sigma^{\prime})
\end{array} \right)\left(\begin{array}{c} U(E,\textbf{r}\sigma^{\prime}) \\
V(E,\textbf{r}\sigma^{\prime})\end{array} \right) =\\
\left(\begin{array}{cc} E+\lambda & 0\\
0 & E-\lambda\end{array}\right)\left(\begin{array}{c} U(E,\textbf{r}\sigma) \\
V(E,\textbf{r}\sigma) \end{array}\right) \,\,
\label{shfb}
\end{split}
\end{eqnarray}

where $\lambda$ is the chemical potential. The local fields $h(\textbf{r},\sigma,\sigma^{\prime})$
and $\Delta(\textbf{r},\sigma,\sigma^{\prime})$ can be calculated in coordinate space. Details can be found in Refs.~\cite{Ring,Stoitsov,Greiner}.

\section{Details of Calculations}
In this work, the ground-state properties of even-even and odd $^{84-140}Mo$ have been reproduced by using the code HFBTHO (v2.00d) \cite{Stoitsov2013} which utilizes the axial Transformed Harmonic Oscillator (THO) single-particle basis to expand quasi-particle wave functions. It iteratively diagonalizes the Hartree-Fock-Bogoliubov Hamiltonian based on generalized Skyrme-like energy densities and zero-range pairing interactions until a self-consistent solution is found.

Calculations were performed with the SLy4 Skyrme functional \cite{Chabanat} as in Ref.~\cite{Stoitsov}, and by using the same parameters as in our previous paper I \cite{Bassem} : A mixed surface-volume pairing with identical pairing strength for both protons and neutrons, and a quasi-particle cutoff of $E_{cut}=60~Mev$. The Harmonic Oscillator basis was characterized by the oscillator length $b_0=-1.0$ which means that the code automatically sets $b_0$ by using the relation $b_0=\sqrt{\hbar / m \omega_0}$, with $\hbar \omega_0=1.2*41/A^{1/3}$. 
The number of oscillator shells taken into account was $N_{max}=16~shells$, the total number of states in the basis $N_{states}=500$, and the value of the deformation $\beta$ is taken from the column $\beta_2$ of the Ref.~\cite{Moller95}. The number of Gauss-Laguerre and Gauss-Hermite quadrature points was $N_{GL} = N_{GH} = 40$, and the number of Gauss-Legendre points for the integration of the Coulomb potential was $N_{Leg} = 80$ \cite{Bassem} .

In the case of odd isotopes, calculations are made by using the blocking of quasi-particle states \cite{Schunck}. The time-reversal symmetry is, by construction, conserved in HFBTHO (v2.00d), the blocking prescription is implemented in the equal filling approximation, and the time-odd fields of the Skyrme functional are identically zero. The identification of the blocking candidate is done using the same technique as in HFODD \cite{Dobaczewski2009} : the mean-field Hamiltonian $h$ is diagonalized at each iteration and provides a set of equivalent single-particle states. Based on the Nilsson quantum numbers of the requested blocked level provided in the input file, the code identifies the index of the quasi-particle (q.p.) to be blocked by looking at the overlap between the q.p. wave-function (both lower and upper component separately) and the s.p. wave-function. The maximum overlap specifies the index of the blocked q.p. \cite{Stoitsov2013}.

The different parameters sets of Skyrme forces for prediction of the nuclear ground-state properties are given in \cite{Bartel,Baran}. SLy4 \cite{Chabanat} parameters set used in this study is shown in Table \ref{table1}.

\begin{table}[ht]
	\centering
	\caption{SLy4 parameters set.\label{table1}}
	\begin{tabular}{@{}c@{\hspace{18pt}}@{\hspace{18pt}}c@{}} \hline
		Parameter & SLy4 \\ \hline
		t$_0$ (MeV fm$^3$)   &     -2488.91 \\
		t$_1$ (MeV fm$^5$)   &       486.82 \\
		t$_2$ (MeV fm$^5$)   &      -546.39 \\
		t$_3$ (MeV fm$^4$)   &       13777.0 \\
		x$_0$ 		    &       0.834 \\
		x$_1$    	    &       -0.344 \\
		x$_2$   	    &      -1.000 \\
		x$_3$  		    &       1.354 \\
		W$_0$ (MeV fm$^3$)   &       123.0	\\
		$\sigma$            &        1/6    \\  \hline
	\end{tabular}
\end{table}

As in paper I, in the input data file of HFBTHO program (v2.00d) \cite{Stoitsov2013}, we have modified the values of the pairing strength for neutrons $V_0^{n}$ and protons $V_0^{p}$ (in MeV), which may be different, but in our study we have used the same pairing strength $V_0^{n,p}$ for both. At each time, we have executed the program and compared the obtained ground-state energy with the experimental value. This procedure was repeated until we found the value of $V_0^{n,p}$ that gives the ground-state energy closest to the experimental one. For more details, see paper I \cite{Bassem}  and references therein.

By fitting the obtained values of $V_0^{n,p}$ to  $A$, we have found the following formula:
\begin{equation}
\large {{ V_0^{n,p} = 155.88\,A^{\frac{1}{6}}}}
\label{eqV0}
\end{equation}

In order to calculate the ground-state properties for both even-even and odd $^{84-140}Mo$ isotopes, the equation (\ref{eqV0}) has been used to generate the pairing-strength $V_0^{n,p}$ that we have included in the code HFBTHO (v2.00d). The same calculations have been performed for $^{88-144}Ru$ isotopes. The results are presented in the next section.

\section{Results and Discussion}
In this section we present the numerical results of this work, particularly for binding energy, double and single neutron separation energies, charge, neutron and proton radii, pairing gap and  quadrupole deformation for $^{84-140}Mo$  and $^{88-144}Ru$ isotopes.\\
In all our calculations, we used the Skyrme force (SLy4) and Eq.(\ref{eqV0}) for the pairing strength.

\subsection{Binding energy}

\begin{figure}[!htb]
	\minipage{0.48\textwidth}
	\centerline{\psfig{file=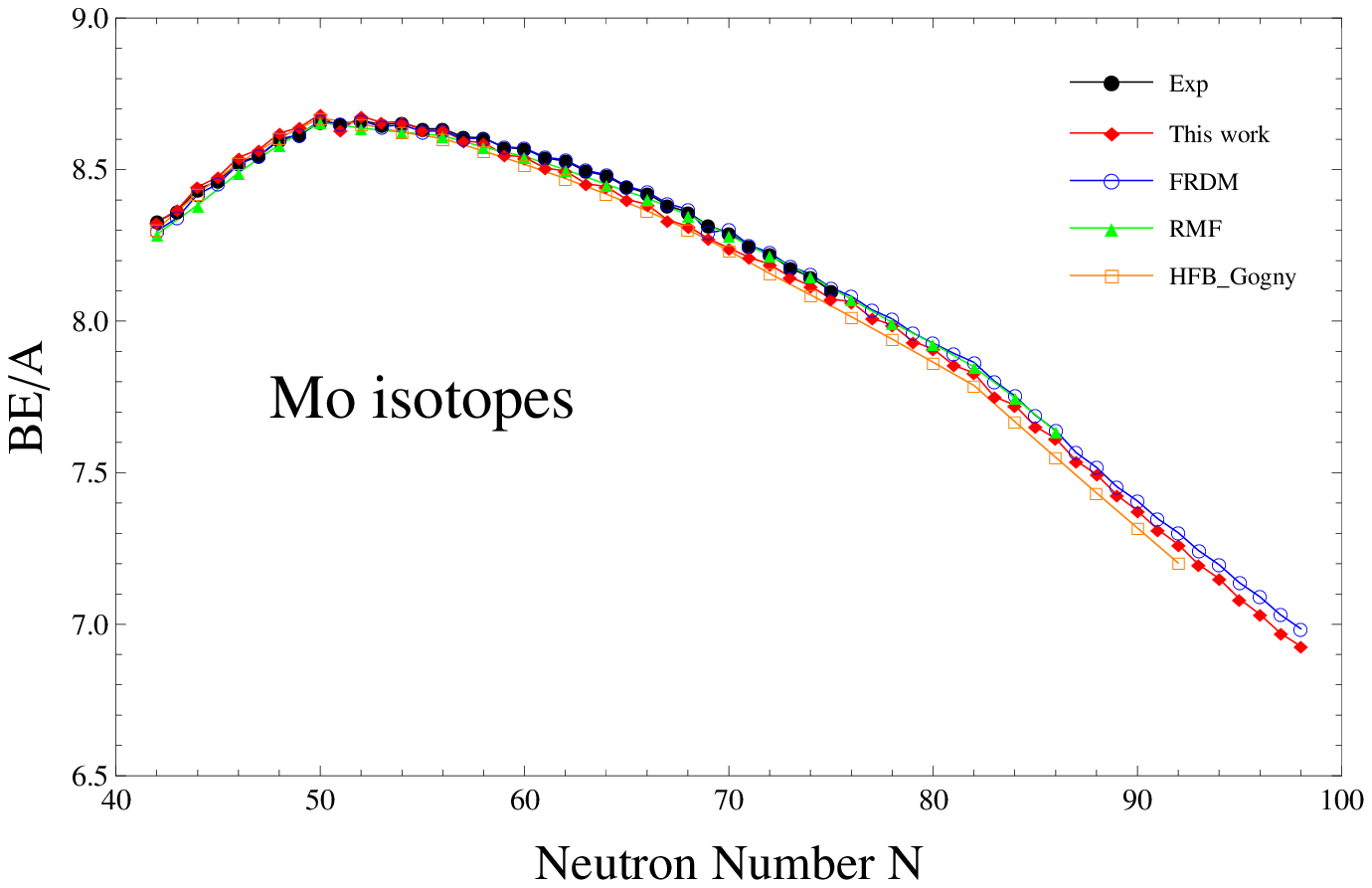,width=\linewidth, height=4cm}}
	\endminipage\hfill
	\minipage{0.48\textwidth}
	\centerline{\psfig{file=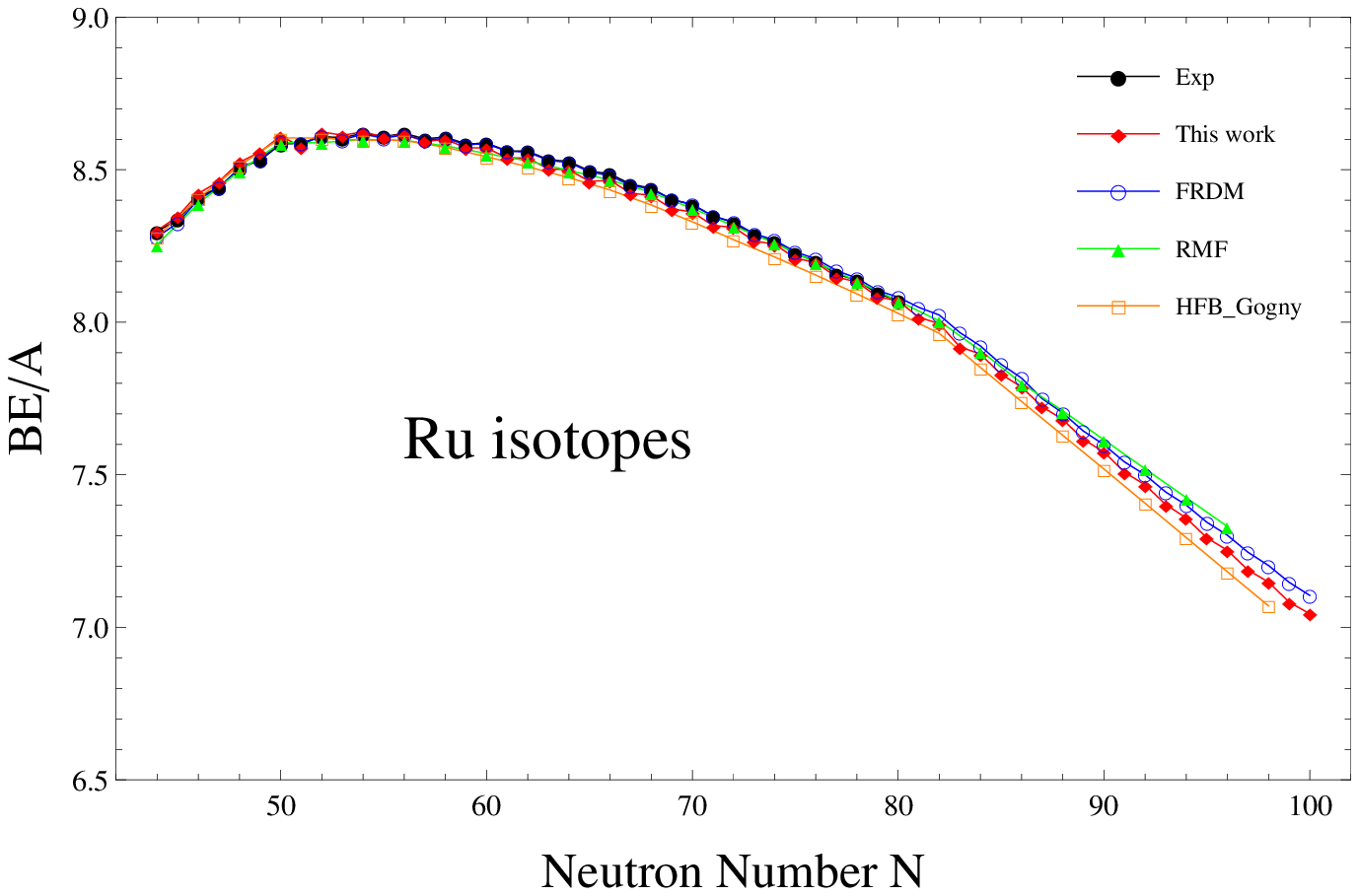,width=\linewidth, height=4cm}}
	\endminipage\hfill
	\caption{(Color online) Binding energies per nucleon for even-even and odd isotopic chains of $Mo$ and $Ru$  nuclei.}
	\label{BEexp}
\end{figure}

The calculated Binding Energies (BE) per nucleon for $Mo$ and $Ru$ isotopes, obtained by using the pairing strength generated by Eq.~(\ref{eqV0}) are plotted as function of the neutron number $N$ in Fig. \ref{BEexp}.
The experimental binding energies per nucleon \cite{WANG}, the HFB calculations based on the D1S Gogny force \cite{AMEDEE}, the predictions of Finite Range Droplet Model (FRDM) \cite{Moller97} and Relativistic Mean Field (RMF) model with NL3 functional \cite{Lalazissis} are shown in Fig. \ref{BEexp} for comparison.

From Fig. \ref{BEexp}, it is seen that the binding  energies per particle for $Mo$ and $Ru$ isotopes  produced by our calculations using HFB with SLy4 parameter set, are in good agreement with the experimental data. We note that the maximums in the binding energy per nuclei, (BE/A), for both $Mo$ and $Ru$ isotopes, are observed at the magic neutron numbers $N = 50$ and  $N = 82$.

The approximately maximal errors of $BE/A$ between the calculated results in the
present study and the experimental data for both $Mo$ and $Ru$ are listed in Table \ref{table2}. The predictions of FRDM \cite{Moller97} and RMF \cite{Lalazissis} theories as well as the HFB calculations based on the D1S Gogny force \cite{AMEDEE} are listed too for comparison.

\begin{table}[!htb]
	\centering
	\caption{The maximal difference error $(BE/A)_{theor}-(BE/A)_{exp}$ (in Mev).\label{table2}}
	{\begin{tabular}{@{}c@{\hspace{18pt}}c@{\hspace{18pt}}c@{\hspace{18pt}}c@{\hspace{18pt}}c@{}} \hline
			Nuclei 		&    This work 	&   RMF		&  FRDM		&	HFB$_{Gogny}$\\ \hline
			Mo		&	0.04772		&	0.05180		&	0.03162 	&	0.06286	\\
			Ru		&	0.03415 	&	0.04323	    &	0.01800	    &	0.05613 \\
			\hline
		\end{tabular}
	}
\end{table}

\begin{figure}[!htb]
	\minipage{0.48\textwidth}
	\centerline{\psfig{file=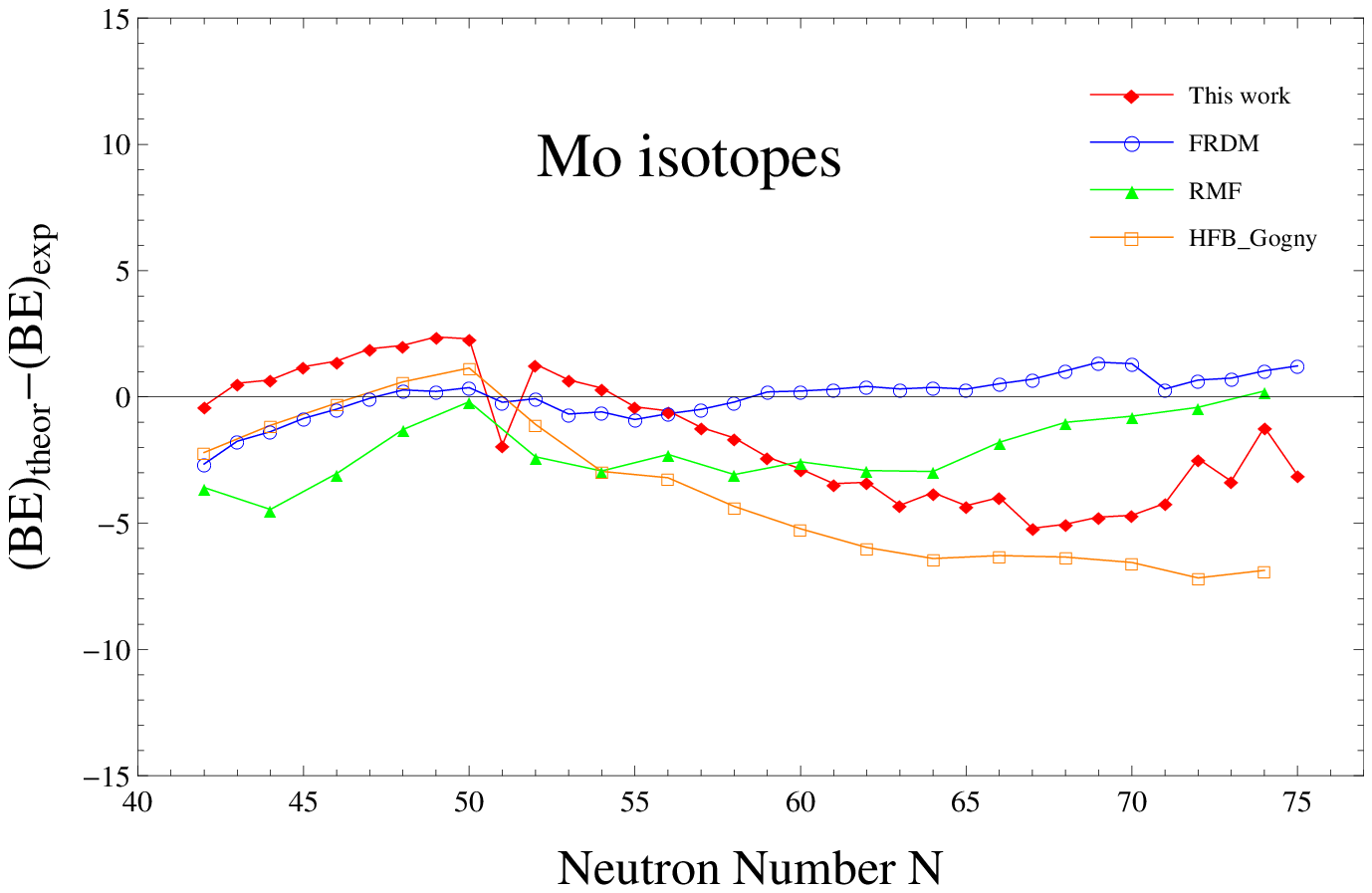,width=\linewidth, height=4cm}}
	\endminipage\hfill
	\minipage{0.48\textwidth}
	\centerline{\psfig{file=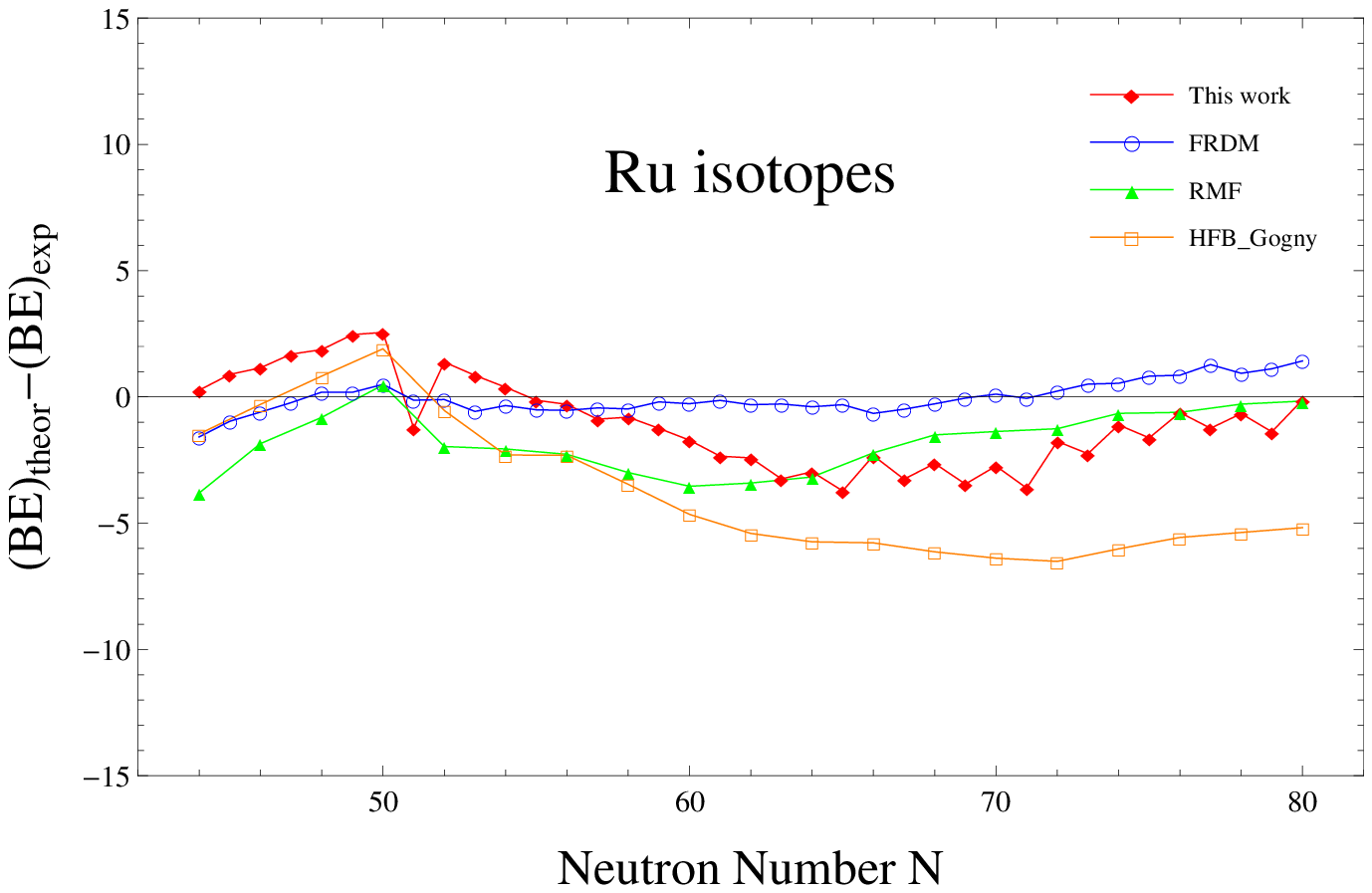,width=\linewidth, height=4cm}}
	\endminipage\hfill
	\caption{(Color online) Binding energies per nucleon for even-even and odd isotopic chains of $Mo$ and $Ru$  nuclei.}
	\label{Delta_BE}
\end{figure}

In order to show to what extent our results are accurate, The differences between the experimental total BE and the calculated results obtained in this work by using Eq.~(\ref{eqV0}) are shown as function of the neutron number $N$ in Fig. \ref{Delta_BE}. The HFB calculations based on the D1S Gogny force \cite{AMEDEE} as well as the predictions of Finite Range Droplet Model (FRDM) \cite{Moller97} and Relativistic Mean Field (RMF) model with NL3 functional \cite{Lalazissis} are also included for comparison. We point out that this comparison is made only for isotopes that have experimental data.

The mean absolute error between experimental data and results of this work for both $Mo$, and $Ru$ are listed in Table \ref{table3}. The predictions of FRDM and RMF theories as well as the HFB calculations based on the D1S Gogny force  are also listed for comparison.

\begin{table}[!htb]
	\centering
	\caption{The mean absolute error $(BE)_{theor}-(BE)_{exp}$ (in Mev).\label{table3}}
	{\begin{tabular}{@{}c@{\hspace{18pt}}c@{\hspace{18pt}}c@{\hspace{18pt}}c@{\hspace{18pt}}c@{}} \hline
			Nuclei 		&    This work 	&   RMF		&  FRDM		&	HFB$_{Gogny}$\\ \hline
			Mo		&	    1.5777 		&		2.0809  &	 0.0295	    &	3.7749  \\
			Ru		&	 	0.9790	    &	 1.7599		&	0.0313  	&	3.7027 \\
			\hline
		\end{tabular}
	}
\end{table}

\subsection{Neutron separation energy}
The one-neutron and two-neutron separation energies are very important in investigating the nuclear shell structure. In the present work, we calculated one- and two-neutron separation energies for $Mo$ and $Ru$ isotopes in SLy4 parametrization with the pairing strength $V_0^{n,p}$ generated by Eq.~(\ref{eqV0}).

The double and single neutron separation energies are defined as :
\begin{equation}
S_{2n}(Z,N)=BE(Z,N)-BE(Z,N-2)
\end{equation}
\begin{equation}
S_{n}(Z,N)=BE(Z,N)-BE(Z,N-1)
\end{equation}
Note that when using these equations, all binding energies must be involved with a positive sign.

\begin{figure}[!htb]
	\minipage{0.48\textwidth}
	\centerline{\psfig{file=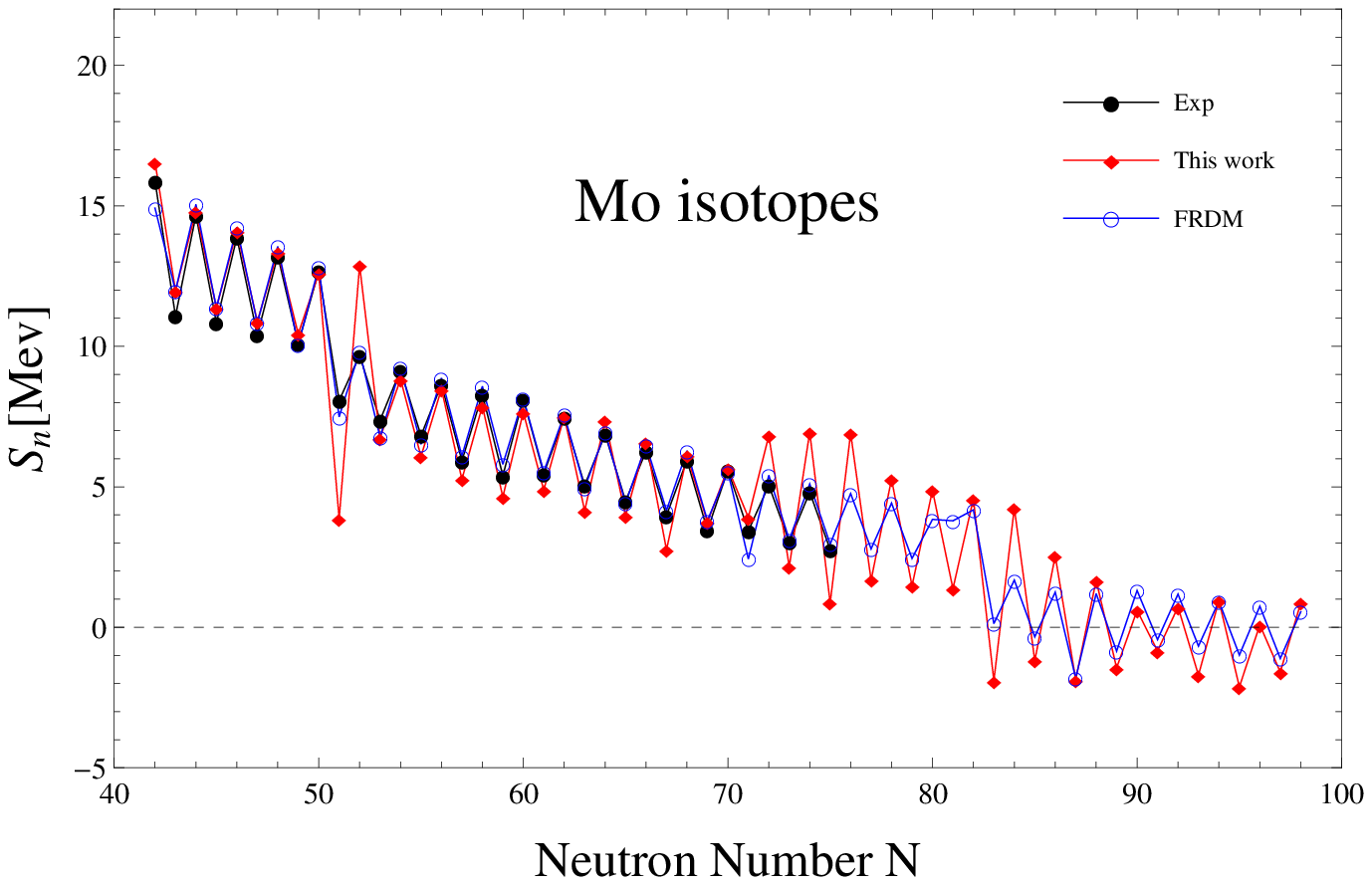,width=\linewidth, height=4cm}}
	\endminipage\hfill
	\minipage{0.48\textwidth}
	\centerline{\psfig{file=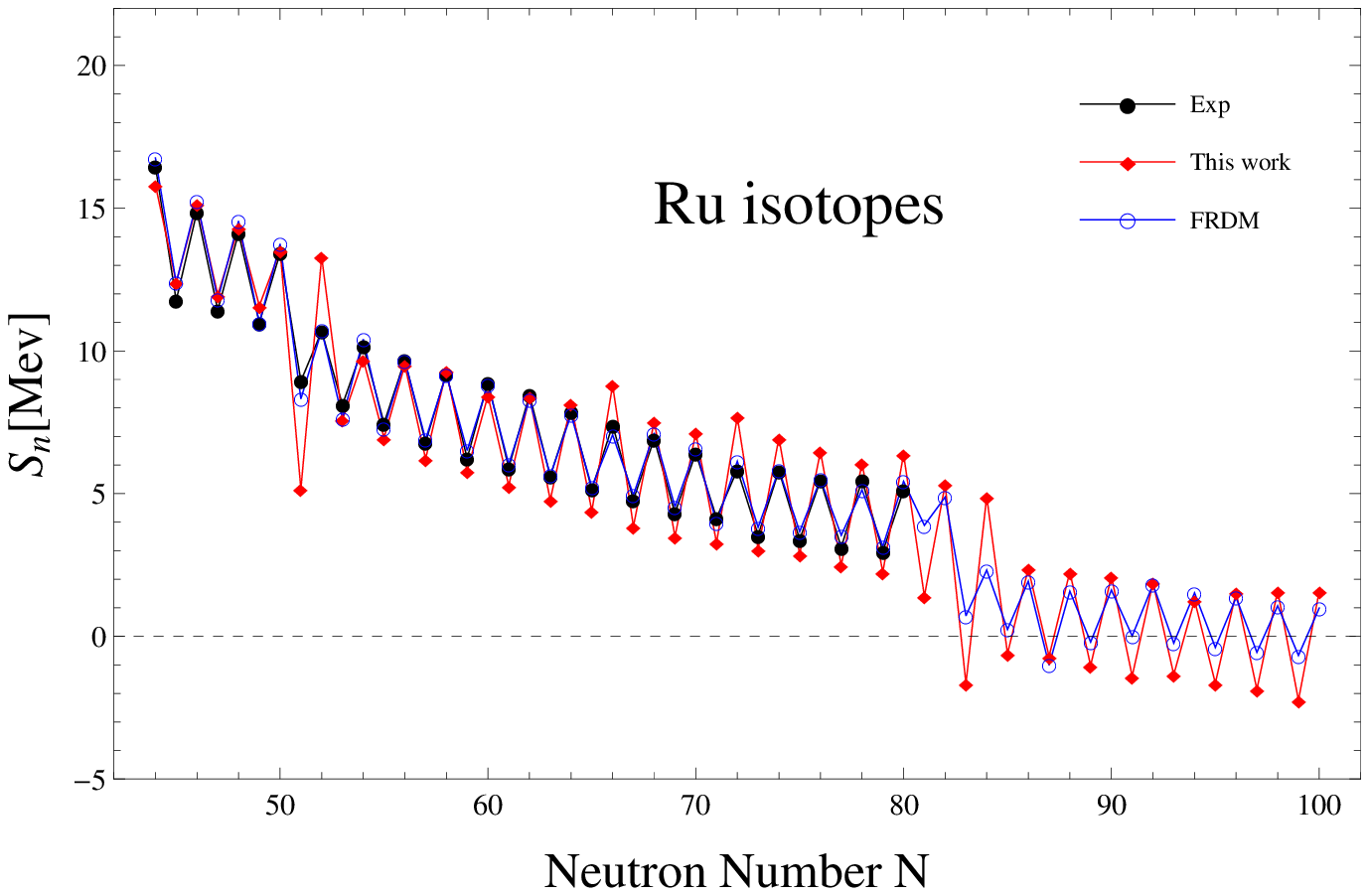,width=\linewidth, height=4cm}}
	\endminipage\hfill
	\caption{(Color online) The  single neutron separation energies, $S_{n}$, of $Mo$ (left), and $Ru$ (right) isotopes.}
	\label{Sn}
\end{figure}

\begin{figure}[!htb]
	\minipage{0.48\textwidth}
	\centerline{\psfig{file=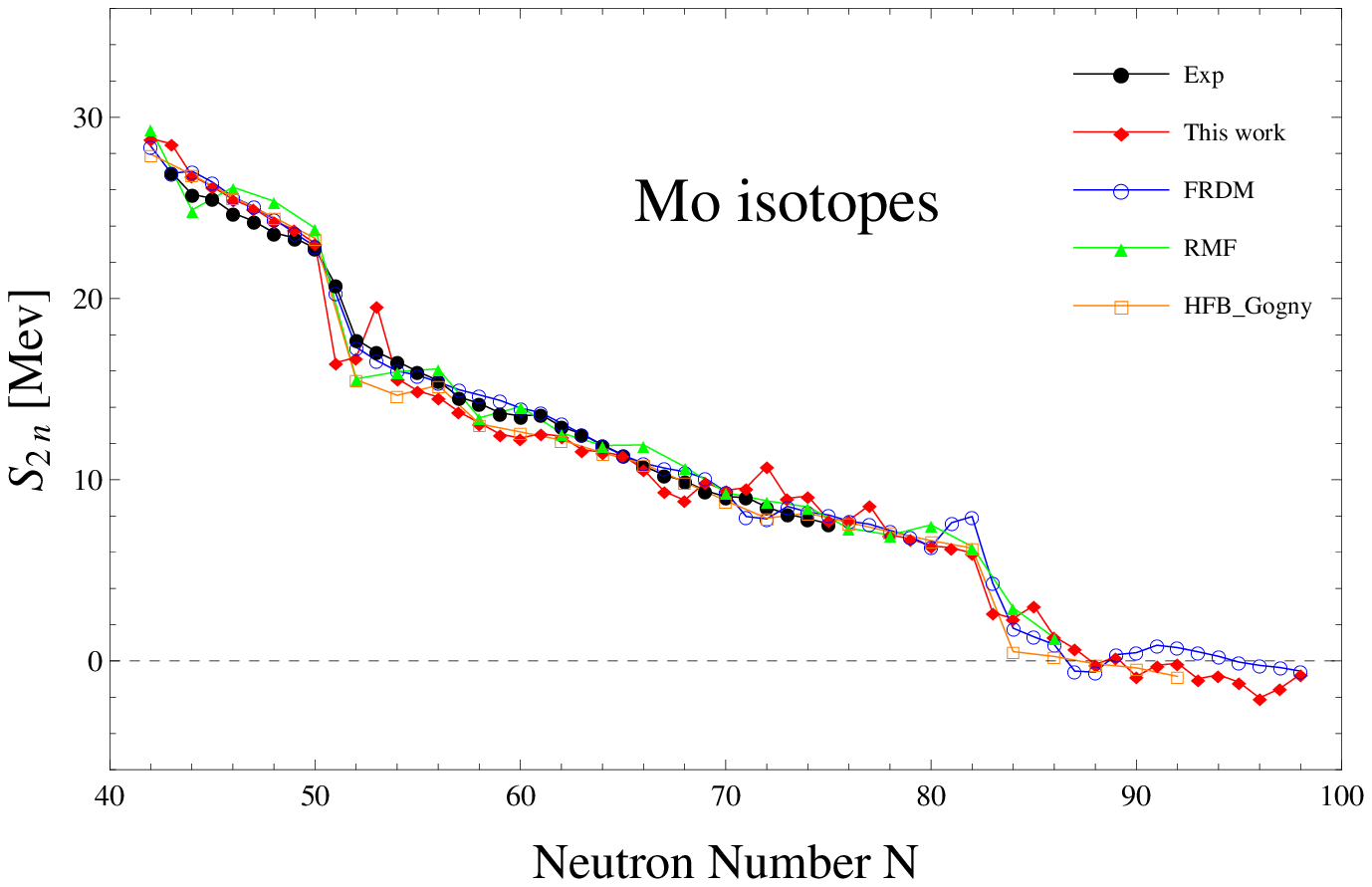,width=\linewidth, height=4cm}}
	\endminipage\hfill
	\minipage{0.48\textwidth}
	\centerline{\psfig{file=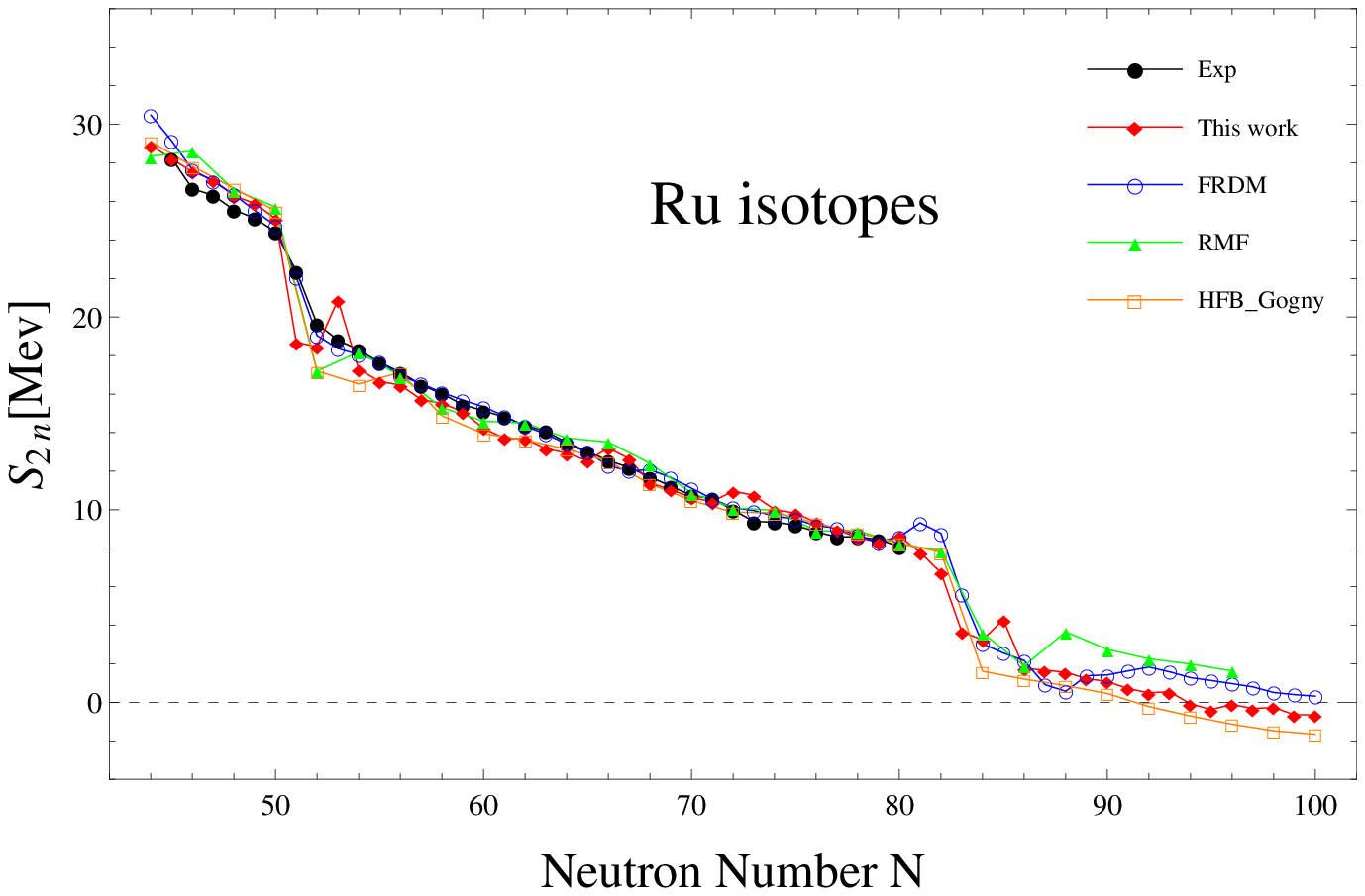,width=\linewidth, height=4cm}}
	\endminipage\hfill
	\caption{(Color online) The  double neutron separation energies, $S_{2n}$, of $Mo$  and $Ru$  isotopes. }
	\label{S2n}
\end{figure}

The calculated $S_{n}$ and $S_{2n}$ for $Mo$ and $Ru$ isotopes  are displayed in Figs. \ref{Sn} and \ref{S2n}, respectively. The available experimental data \cite{WANG}, HFB calculations based on the D1S Gogny force \cite{AMEDEE} and predictions of RMF \cite{Lalazissis} and FRDM \cite{Moller97} theories are presented for comparison.\\

It is clearly seen from Figs. \ref{Sn} and \ref{S2n} that the experimental separation energies are reproduced almost well by our calculations. Even if there are slight differences in the case of some isotopes, our results have the lowest absolute mean error in comparison with the other theories, as we can see from Table \ref{table4}. Also, sharp decreases in $S_{n}$ and $S_{2n}$ are seen at the magic neutron numbers $N=50$ and  $N=82$, which correspond to  closed shells. 

\begin{table}[!htb]
	\centering
	\caption{The mean absolute error $(S_{2n})_{theor}-(S_{2n})_{exp}$  and  $(S_{n})_{theor}-(S_{n})_{exp}$ (in Mev).\label{table4}}
	\begin{tabular}{c|c|c|c|c||c|c|}
		\cline{2-7}
		& \multicolumn{4}{c||}{ $(S_{2n})_{theor}-(S_{2n})_{exp}$ } & \multicolumn{2}{c|}{$(S_{n})_{theor}-(S_{n})_{exp}$} \\ \cline{2-7} 
		&  This work   &  RMF   &  FRDM   &  HFB$_{Gogny}$   &    This work       &  FRDM         \\ \hline
		\multicolumn{1}{|c|}{Mo} &  0.090   & 0.240  & 0.200 &  0.290 &  0.061   &  0.086
		         \\ \hline
		\multicolumn{1}{|c|}{Ru} &  0.075  &  0.202   & 0.166    &  0.206  &  0.028         &   0.089    \\ \hline  
	\end{tabular}
\end{table}

For nuclei with $N>82$, both the double and single neutron separation energies are quite small, and  $S_{n}$ of these nuclei with odd mass numbers are almost zero or negative. The two neutron drip-line nucleus is predicted by our calculations to be $^{132}Mo$ and $^{138}Ru$ for $Mo$ and $Ru$ isotopes, respectively. 
Another well-known characteristic shown in Fig. \ref{Sn} is that even nuclei have  $S_{n}$ greater than their odd neighbors, this is what explains the fluctuations seen in Fig. \ref{Sn}.

\subsection{Neutron, Proton and Charge radii}

\begin{figure}[!htb]
	\minipage{0.48\textwidth}
	\centerline{\psfig{file=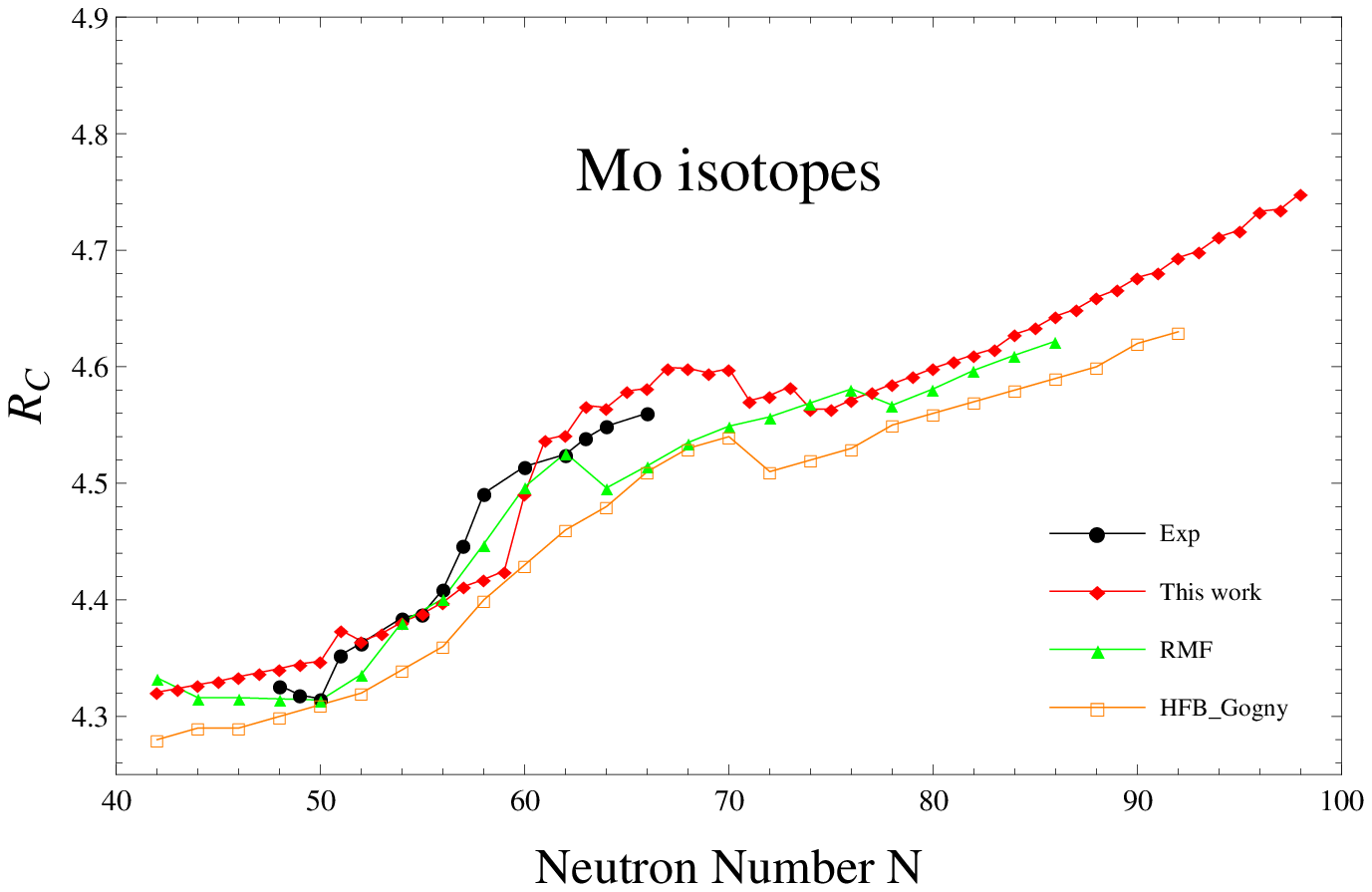,width=\linewidth, height=4cm}}
	\endminipage\hfill
	\minipage{0.48\textwidth}
	\centerline{\psfig{file=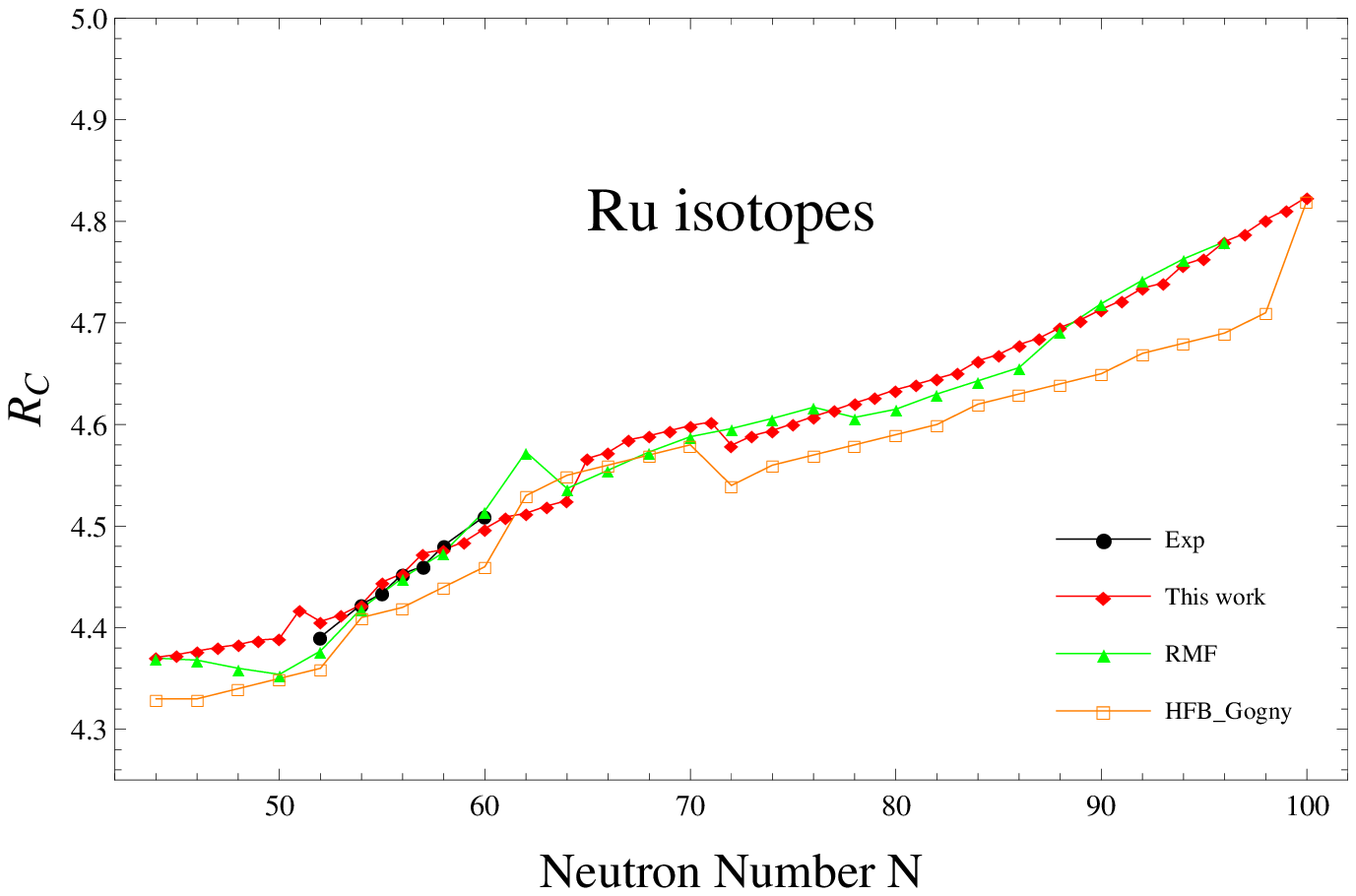,width=\linewidth, height=4cm}}
	\endminipage\hfill
	\caption{(Color online) The charge radii of $Mo$ and $Ru$ isotopes.}
	\label{Rc}
\end{figure}

In Fig. \ref{Rc}, the root mean square charge radii, $R_c$, (which are calculated using  $R_c^2=R_p^2+0.64~(fm)$, where $R_c$ and $R_p$ are charge and proton radii, respectively) are shown. The available experimental data \cite{Angeli}, the predictions of RMF theory \cite{Lalazissis} and the HFB calculations based on the D1S Gogny force \cite{AMEDEE} are also shown in  Fig .\ref{Rc} for comparison.

From Fig. \ref{Rc}, a good agreement between theory and experiment can be clearly seen.
One can also see that the charge radii for nuclei heavier than the closed neutron shell, $N=50$, start increasing with addition of neutrons.

\begin{figure}[!htb]
	\minipage{0.48\textwidth}
	\centerline{\psfig{file=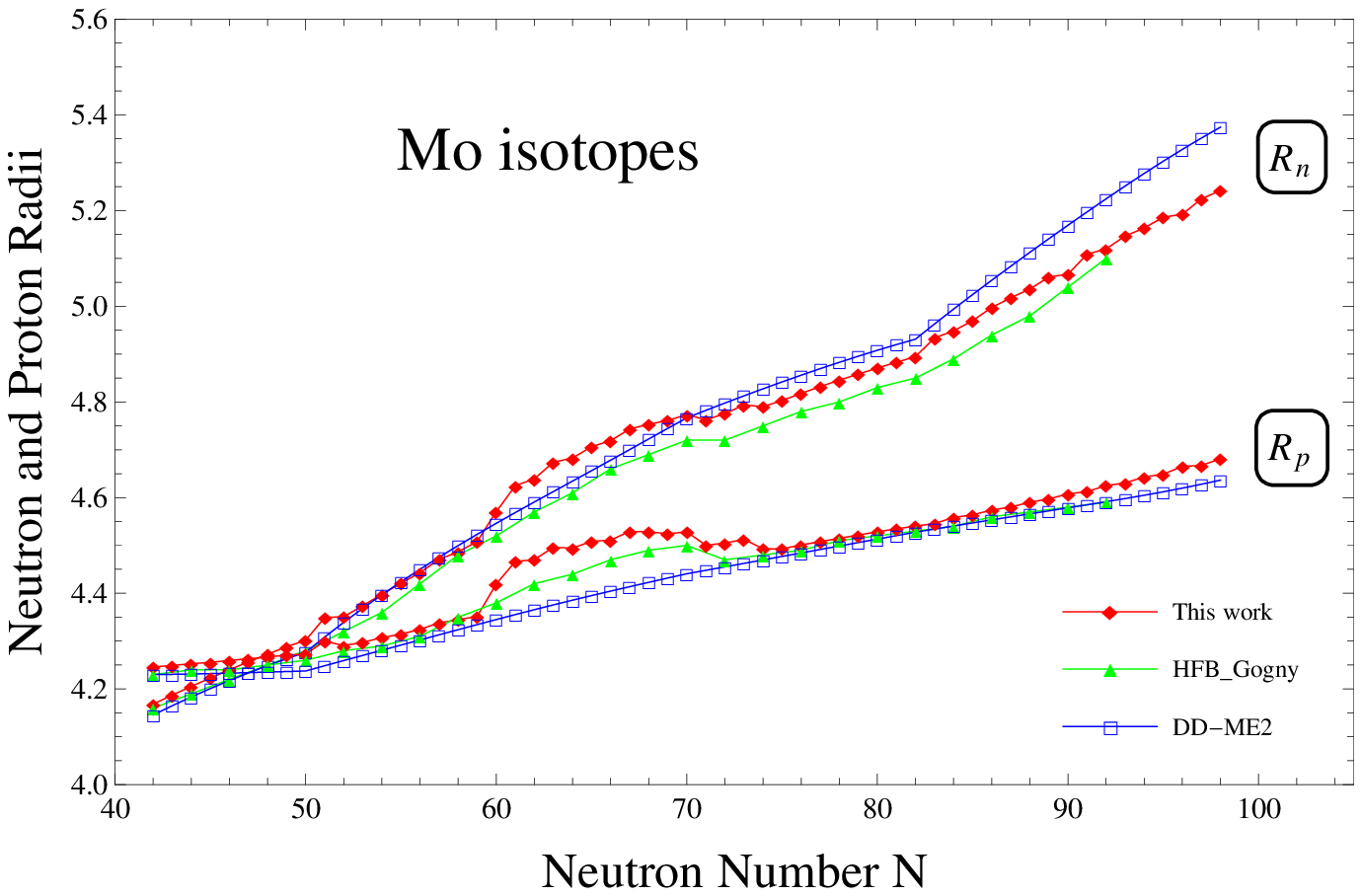,width=\linewidth, height=4cm}}
	\endminipage\hfill
	\minipage{0.48\textwidth}
	\centerline{\psfig{file=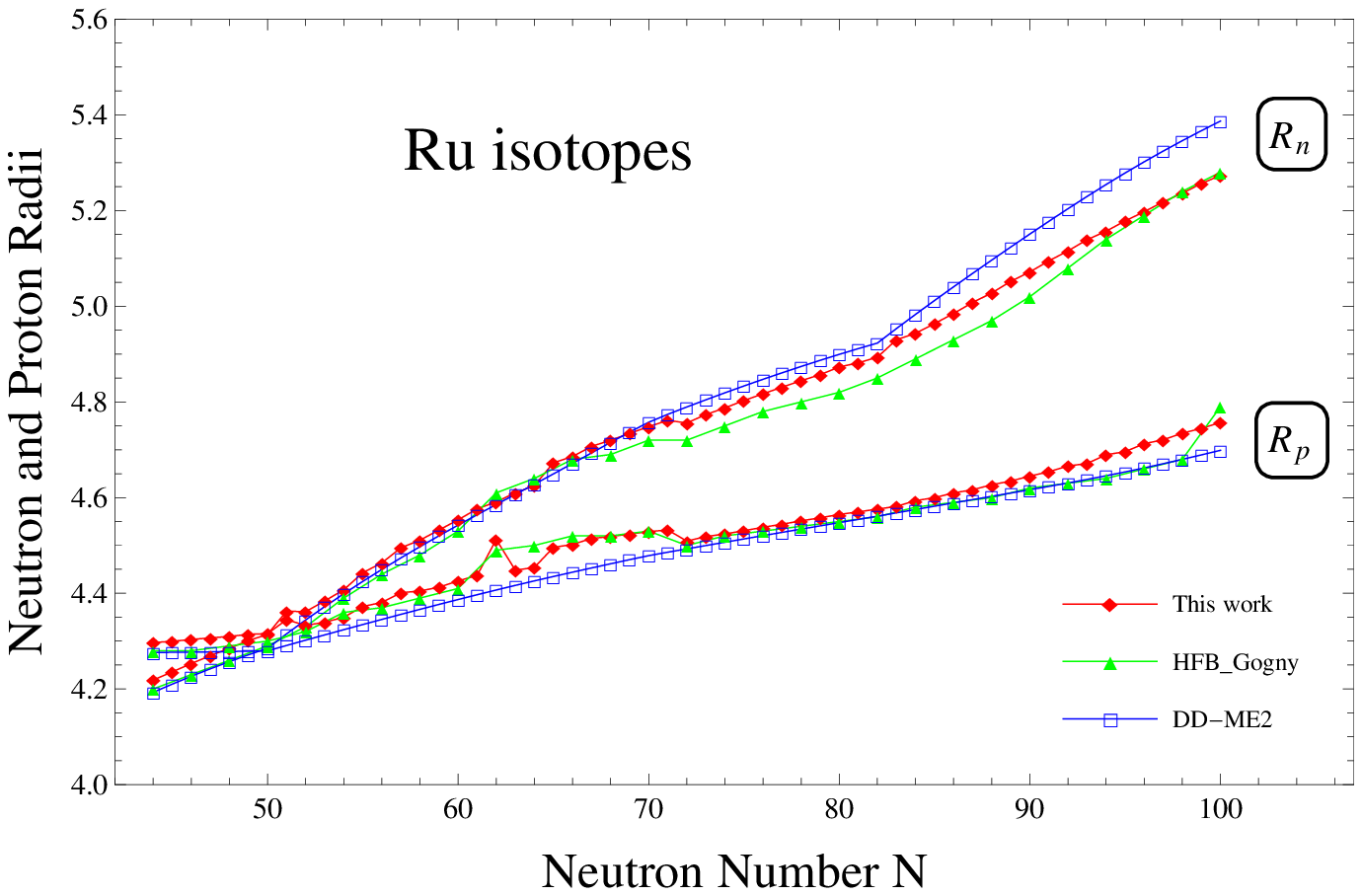,width=\linewidth, height=4cm}}
	\endminipage\hfill
	\caption{(Color online) The neutron and proton radii of $Mo$ and $Ru$ isotopes.}
	\label{R}
\end{figure}

Fig. \ref{R} shows the neutron and proton radii of $Mo$ and $Ru$ isotopes obtained in our calculations. HFB calculations based on the D1S Gogny force \cite{AMEDEE} are shown for comparison as well as results of the relativistic Hartree-Bogoliubov (RHB) model with the DD-ME2 effective interaction calculated by using the code DIRHBZ \cite{code}. We have plotted neutron and proton radii ($R_n$ and $R_p$) together in order to see the difference between them.

As it can be seen from Fig. \ref{R}, the difference between the neutron and proton rms radii starts to increase with the increase of the neutron number, in favor of developing a neutron skin. This difference reaches $0.560$ fm for $^{140}$Mo and $0.516$ fm for $^{144}$Ru, which can be considered as an indication of possible neutron halo in $Mo$ and $Ru$ isotopes. But near the $\beta$-stability line ($N \approx  Z$), the neutron and proton radii are almost the same.  

\subsection{Neutron pairing gap}

The pairing gap is not directly accessible in experiments. Therefore, there are various finite-difference formulas in the literature, which are often interpreted as a measure of the empirical pairing
gap, such as :

The three-point difference formula \cite{satula1} :
\begin{equation}
\Delta_Z^{(3)}(N):=\frac{\pi_{N}}{2}\large[E_b(Z,N-1)-2 \, E_b(Z,N) + E_b(Z,N+1)]
\label{delta3}
\end{equation}
where $N$ and $Z$ are the neutron and proton numbers and $E_b$ is the binding energy (negative) of the nucleus. $\pi_N=(-1)^N$ is the number parity.

Another commonly used relation is the four-point difference formula \cite{Krieger,Cwiok,Bohr98} :
\begin{equation}
\Delta_Z^{(4)}(N):=\frac{\pi_{N}}{4}\large[E_b(Z,N-2)-3 \, E_b(Z,N-1) + 3 \, E_b(Z,N) - E_b(Z,N+1)]
\label{delta4}
\end{equation}

The next order corresponds to the five-point difference formula \cite{Bender2000} :
\begin{equation}
\begin{split}
\Delta_Z^{(5)}(N):=-\frac{\pi_{N}}{8}\large[E_b(Z,N+2) & -4 \, E_b(Z, N+1) + 6 \, E_b(Z,N) \\ &- 4 \, E_b(Z,N-1) + E_b(Z,N-2)]
\end{split}
\label{delta5}
\end{equation}

\begin{figure}[!htb]
	\minipage{0.48\textwidth}
	\centerline{\psfig{file=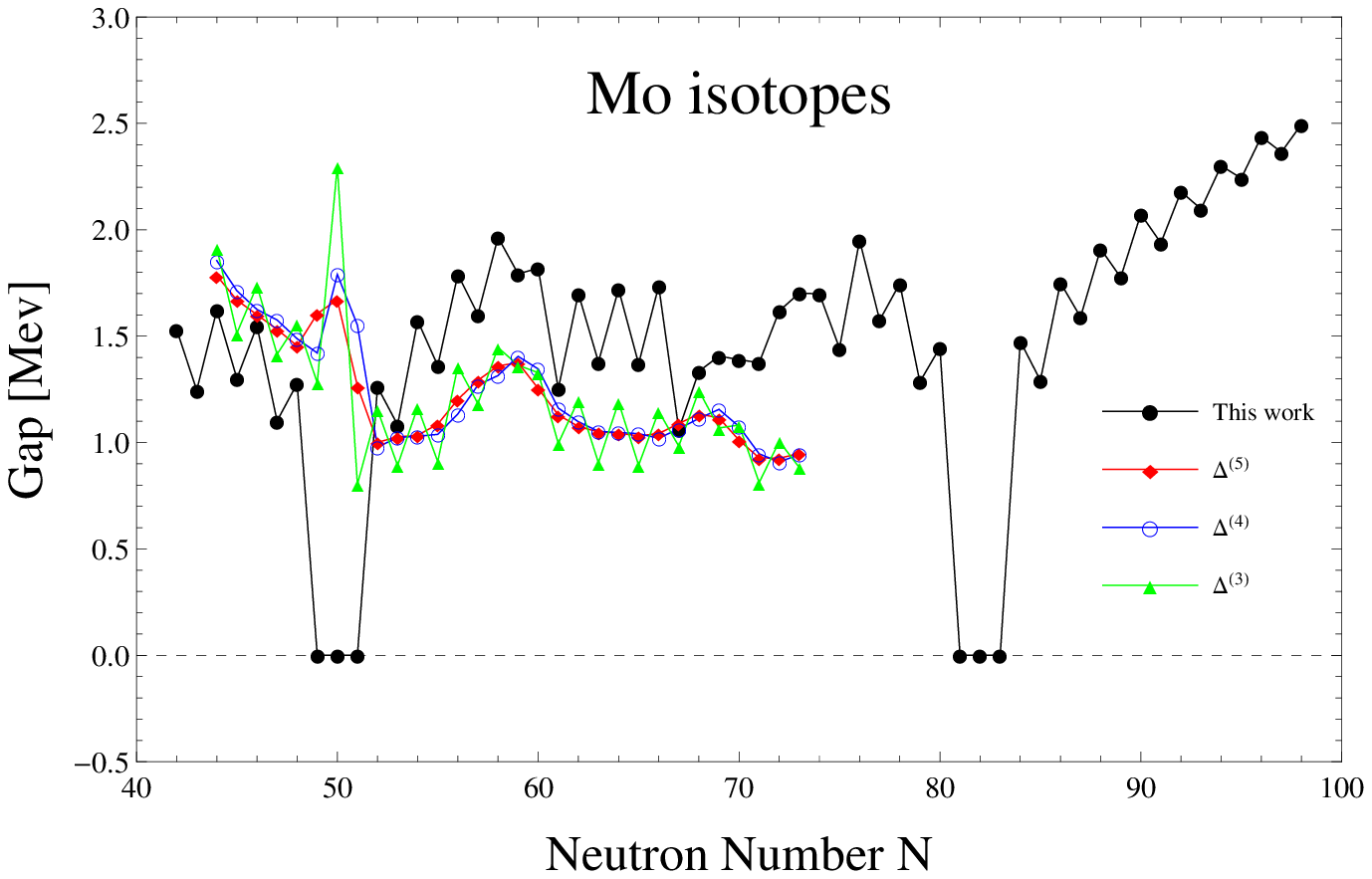,width=\linewidth, height=4cm}}
	\endminipage\hfill
	\minipage{0.48\textwidth}
	\centerline{\psfig{file=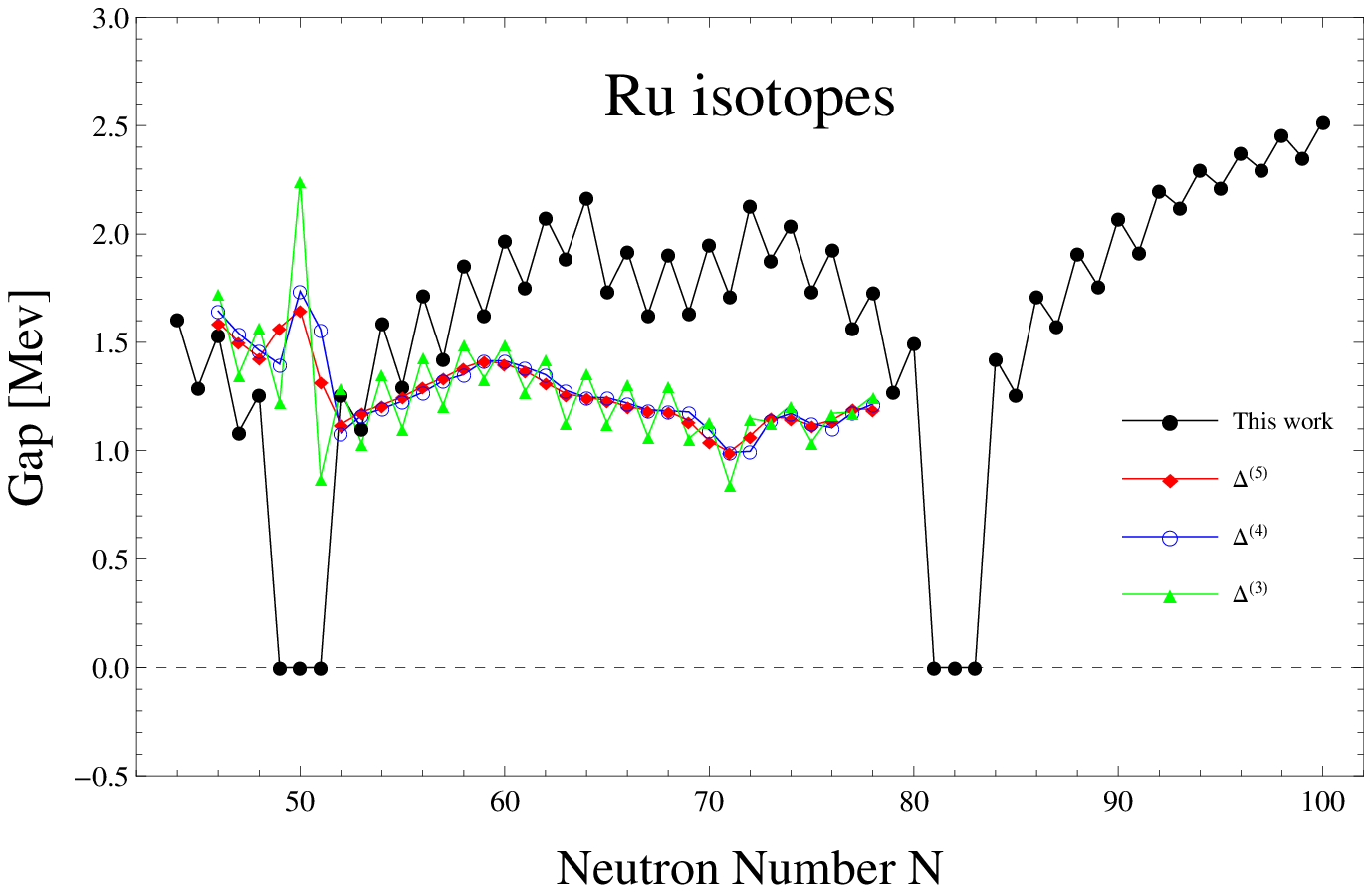,width=\linewidth, height=4cm}}
	\endminipage\hfill
	\caption{(Color online) The neutron pairing gaps of $Mo$ and $Ru$ isotopes.}
	\label{gap}
\end{figure}

In Fig. \ref{gap}, the neutron pairing gaps obtained in our HFB calculations are compared to the experimental data as it has been done in Ref. \cite{Bender} with spectral gaps $\textless uv\Delta \textgreater $. The experimental values
are calculated from the binding energies given in Ref. \cite{WANG} by using the three-point  $\Delta^{(3)}$, the four-point  $\Delta^{(4)}$ and the five-point $\Delta^{(5)}$ formulas, respectively.

From Fig. \ref{gap}, it can be clearly seen the overall closeness
of our calculations with experimental data. The differences between our results and the experimental data are due to the fact that the pairing gaps obtained in our HFB calculations are effective gaps defined in HFBTHO as the mean value of the pairing field 
\begin{equation}
\bar{\Delta}=\frac{Tr(\Delta \, \rho)}{Tr(\rho)},
\end{equation}
with $\rho$ is the normal one-body density matrix and $\Delta$ is the pairing field \cite{Stoitsov}. Such  defined effective gaps have the same behavior as the spectral gaps in Ref. \cite{Bender}.
 
Also, our calculated gaps are exactly zero for closed shell nuclei $N=50$, $N=82$ and their adjacent odd-mass number nuclei as in Figure 4 of the Ref. \cite{Bender}. This is because, at the HFB approximation, the pairing correlations automatically vanish in the case of magic nuclei. The pairing field becomes zero, so the average value of this field vanishes also, which corresponds to a maximum value of the gaps given by the empirical formulas \ref{delta3}, \ref{delta4} and \ref{delta5}.
 
Another remark that can be seen in this figure is that even-even nuclei have larger gaps than odd-mass nuclei, this is due to the blocking of one state in the odd-mass nucleus. The blocked state does not contribute to the pairing potential, leading to overall smaller single particle gaps \cite{Bender2000}.

\subsection{Quadrupole deformation}  
The deformation is an extremely important property  for nuclei, it plays a crucial role in determining their properties such as quadrupole moment, nuclear sizes and isotope shifts. The deformation can also make a deformed shape more favored than the spherical one for certain nuclei by increasing the nuclear binding energy.

\begin{figure}[!htb]
	\minipage{0.48\textwidth}
	\centerline{\psfig{file=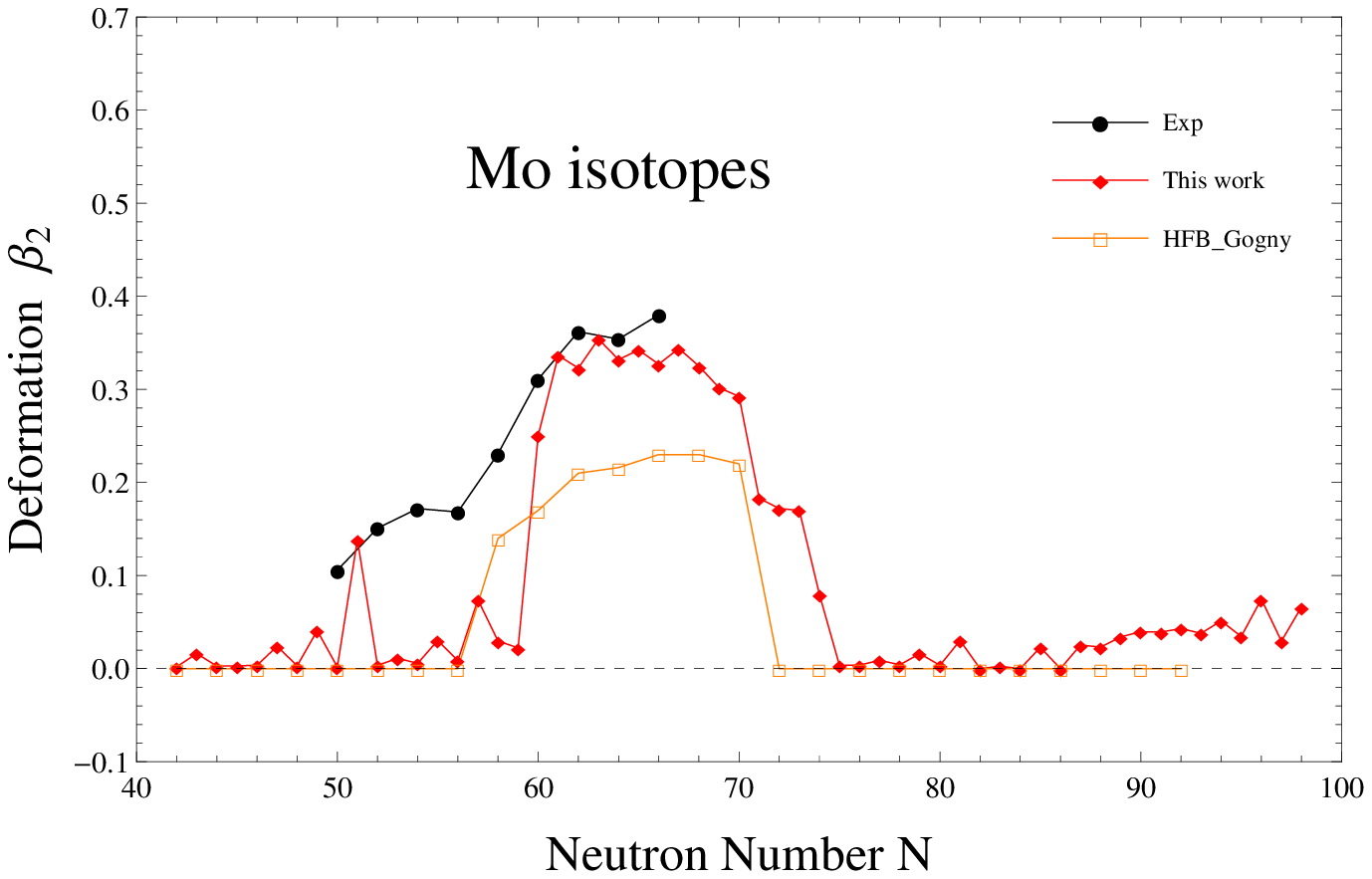,width=\linewidth, height=4cm}}
	\endminipage\hfill
	\minipage{0.48\textwidth}
	\centerline{\psfig{file=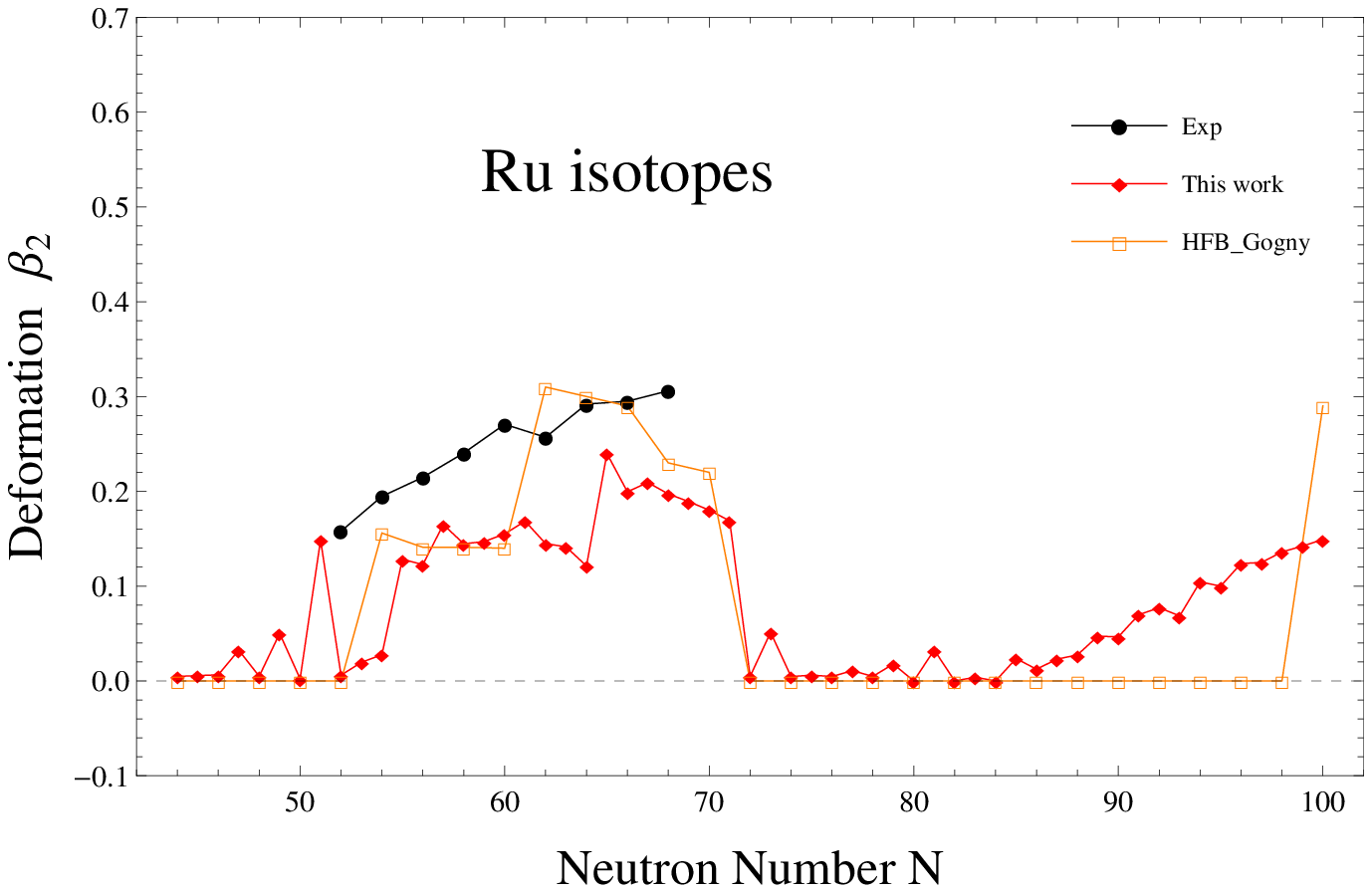,width=\linewidth, height=4cm}}
	\endminipage\hfill
	\caption{(Color online) The quadrupole deformation parameters,  $\beta_2$, for $Mo$ and $Ru$ isotopes.}
	\label{beta}
\end{figure}

In Fig. \ref{beta}, we plot the quadrupole deformation parameter, $\beta_2$, for $Mo$ and $Ru$ isotopes. The results of our calculations are compared with the available experimental data \cite{beta_exp} and results of  HFB calculations based on the D1S Gogny force \cite{AMEDEE}. It must be stressed that Fig. \ref{beta} does not indicate  the sign of the quadrupole deformation parameter $\beta_2$.

As it can be seen from Fig. \ref{beta}, the agreement between our calculations and the experimental data is quite good in general. The $\beta_2$ values show  minima at the magic neutron numbers $N=50$ and $N=82$ as expected, because nearly all nuclei
with $N=50$ or $N=82$ are spherical. Nuclei below and above these two magic neutron numbers show an interesting change of shape. For nuclei above $N=50$, the prolate deformation increases and then saturates at a value close to $\beta_2 \approx 0.3$ for both $Mo$ and $Ru$ isotopic chains, and for nuclei below $N=82$, there is a transition from deformed to spherical shape. Therefore, the regions where nuclei are moderately deformed are $60 \leqslant N \leqslant 74$ and $55 \leqslant N \leqslant 70$ for $Mo$ and $Ru$, respectively.

\section{Conclusion}
In this work, we have studied two of the most interesting isotopic chains in the periodic table, $Mo$ and $Ru$, for a wide range of neutron numbers. Calculations have been performed by  using HFB method with SLy4 Skyrme force and a new generalized formula for pairing strength $V_0^{n,p}$ for neutrons and protons. The results of these calculations reproduce the available experimental data very well including binding energy per nucleon, the one- and two-neutron separation energies, proton, neutron and charge radii, neutron pairing gap and quadrupole deformation. The parabolic behavior of the BE/A has been  well reproduced in respect to the experimental curve. A possible neutron halo has been observed in both $Mo$ and $Ru$ isotopes. The last stable nucleus against neutron emission was found to be $^{132}Mo$ for $Mo$ isotopes, and $^{138}Ru$ for  $Ru$ isotopes. Indications on shape phase transition for both isotopic chains were given.

\end{document}